\documentclass[12pt]{iopart}

\usepackage{graphicx}
\newcommand{\be}{\begin{equation}} 
\newcommand{\ee}{\end{equation}}
\newcommand{\bea}{\begin{eqnarray}} 
\newcommand{\eea}{\end{eqnarray}}
\begin{document}

\title{Improved grand canonical sampling of vapour-liquid transitions}
\author{Nigel B. Wilding} 
\address{Department of Physics, University of Bath, Bath BA2 7AY,
United Kingdom} 

\begin{abstract}

Simulation within the grand canonical ensemble is the method of choice
for accurate studies of first order vapour-liquid phase transitions in
model fluids. Such simulations typically employ sampling that is
biased with respect to the overall number density in order to overcome
the free energy barrier associated with mixed phase states. However,
at low temperature and for large system size, this approach suffers a
drastic slowing down in sampling efficiency. The culprits are
geometrically induced transitions (stemming from the periodic boundary
conditions) which involve changes in droplet shape from sphere to
cylinder and cylinder to slab. Since the overall number density
doesn't discriminate sufficiently between these shapes, it fails as an
order parameter for biasing through the transitions. Here we report
two approaches to ameliorating these difficulties. The first
introduces a droplet shape based order parameter that generates a
transition path from vapour to slab states for which spherical and
cylindrical droplet are suppressed. The second simply biases with
respect to the number density in a tetragonal subvolume of the
system. Compared to the standard approach, both methods offer improved
sampling, allowing estimates of coexistence parameters and
vapor-liquid surface tension for larger system sizes and lower
temperatures.

\end{abstract}

\maketitle
\section{Introduction}
\label{sec:intro}

\begin{figure}[h]
\centering                                      
\includegraphics[type=pdf,ext=.pdf,read=.pdf,width=0.8\columnwidth,clip=true]{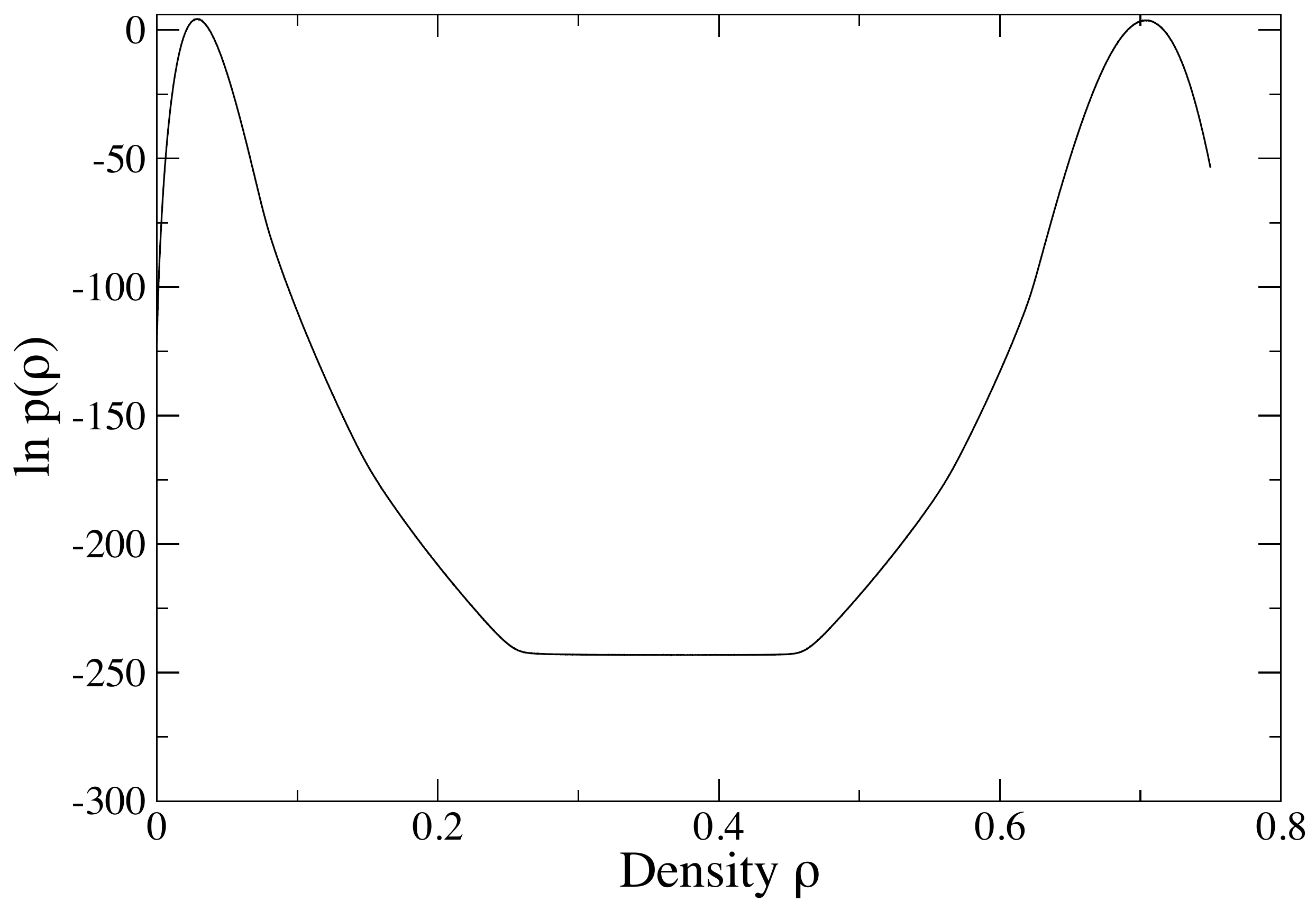}
                                                                                                                                      
\caption{Estimates of the logarithm of the number density
  distribution $p(\rho)$ for the 3d LJ fluid at $T=0.775 T_c$. The
  system volume is $V=(17.5\sigma)^3$. Note the equality of the peak
  weights and the deep probability valley, which is flat in the region
  for which liquid slab states occur. The distribution was obtained
  using MUCA biased with respect to the density $\rho$.}
                                                                                                                                      
 \label{fig:prho}
 \end{figure}

An order parameter for a discontinuous phase transition is any
quantity which distinguishes clearly between the states that coexist
on the phase boundary. For magnets this is usually taken to be the
magnetisation, while for the vapour-liquid (VL) transition of a single
component fluid an appropriate choice is the overall number density
$\rho=N/V$.  A particularly successful approach for studying VL phase
transitions utilizes grand canonical Monte Carlo (GCMC) simulation to
study the probability distribution $p(\rho)$ of the fluctuating
density \cite{wilding1995}.  Precisely at coexistence, $p(\rho)$
exhibits a pair of equally weighted peaks which are centred on the
densities of the coexisting phases. These peaks are separated by a
deep probability valley as is illustrated in Fig.~\ref{fig:prho} which
displays the coexistence form of $\ln p(\rho)$ for the three
dimensional (3d) Lennard-Jones (LJ) fluid at a strongly subcritical
temperature. For sufficiently large system size and/or sufficiently
low temperature, the valley exhibits a flat linear region corresponding to
liquid slab configurations in which the liquid spans the periodic
boundaries in $d-1$ dimensions and is separated from the vapor by a
planar interface in the remaining dimension. Since the free energy is
related to the probability distribution by $\beta f(\rho)=-\ln
p(\rho)$, the valley flatness reflects the equality of the free energy
of the coexisting phases, there being no cost involved in altering the
width of the liquid slab. 

Slab configurations differ in free energy
from a pure phase vapor (or a liquid) by the free energy cost of
inserting a pair of vapor-liquid interfaces. Accordingly, the
probability ratio of pure phase and slab states provides an accurate
means of measuring the vapor-liquid surface tension
\cite{Binder:1982ck,Errington2003}. Specifically one finds for a 3d
system of volume $V=L^3$,

\be 
\gamma_{vl}=(2\beta L^2)^{-1}\ln(p_{\rm max}/p_{\rm min})\:,  
\label{eq:gamlv}
\ee 
where $\beta=(k_BT)^{-1}$ and $p_{max}\equiv p_{\rm vap}=p_{\rm liq}$ is the height of the peaks of $p(\rho)$.

Thus, in principle at least,  GCMC estimates of $p(\rho)$ provide a route to determining both coexistence parameters (via the equality of  peak weights or the flatness of the valley bottom) and vapor-liquid surface tension (via the peak to valley height). However the probability valley separating the two pure phases represents a serious obstacle to efficient MC sampling of the distribution. For modest system sizes this can be overcome using Multicanonical sampling (MUCA).  MUCA is a biasing scheme for first order phase transitions which was first introduced in the context of lattice based spin models \cite{Berg1992}. It was subsequently generalized to GCMC studies of fluid phase transitions \cite{wilding1995} and used to study a wide variety of fluid models including some originally introduced by George Stell \cite{pini2001,magee2002,Gibson06}. 

To recall how MUCA works with GCMC, consider a fluid system specified
by a set of $N$ particles and their coordinates ${\bf r}^N$.  MUCA samples not from the true Hamiltonian ${\cal H}({\bf r}^N)$ but from a modified one

\be
{\cal H}^{\prime}({\bf r}^N)={\cal H}({\bf r}^N)+\eta({\bf r}^N)\:,
\ee
where $\eta({\bf r}^N)$ is a user defined weight function. In the
original GCMC formulation \cite{wilding1995}, the weights $\eta({\bf
  r}^N)$ were simply defined with respect to the overall density
ie. $\eta({\bf r}^N)=\eta(\rho)$.  It is straightforward to show
\cite{Bruce2003} that the choice $\eta(\rho)=-f(\rho)$ leads to uniform sampling of the density range.  Of
course, generally one doesn't know the form of $f(\rho)$ {\em
  a-priori}. Nevertheless iterative \cite{Berg96,Virnau:2004bh} or
transition matrix methods \cite{Smith:1995kx} allow the determination
of a good approximate form which should result in near-uniform
sampling. Once such a preweighted simulation has been performed, it is
a simple matter to unfold the bias \cite{Berg1992,Bruce2003} to
determine the requisite form of $p(\rho)$. GCMC in conjunction with density-weighted MUCA  has been successful in yielding the form of $p(\rho)$ at VL phase transitions for moderate system sizes  \cite{wilding1995,Errington2003}. However, for large system sizes and particularly at low temperatures, further sampling difficulties arise due to so-called droplet transitions for which $\rho$ does not provide a good order parameter for biasing purposes, as we now describe. 

\section{Droplet transitions and sampling issues}

Droplet transitions are geometrically induced first order phase
transitions which occur in systems that have periodic boundary
conditions. They were first recorded in the context of the 2D Ising
model \cite{Leung:1990cy}. For 3d fluids and for sufficiently system
volumes, three droplet states are observed. As the density is raised
above that of a homogeneous vapour, a state with a spherical droplet
occurs. Further increasing the density results in a cylindrical
droplet whose ends are connected via the periodic boundary conditions
of the simulation box. Finally, at larger densities a tetragonal slab
configuration is formed. At even higher densities than those at which
slabs occur, a set of analogous `bubble' transitions occur involving
spherical and cylindrical bubbles
\protect\cite{MacDowell:2006a,ARCHER07}.

\begin{figure}[h]
    \centering                               
\includegraphics[type=pdf,ext=.pdf,read=.pdf,width=0.8\columnwidth,clip=true]{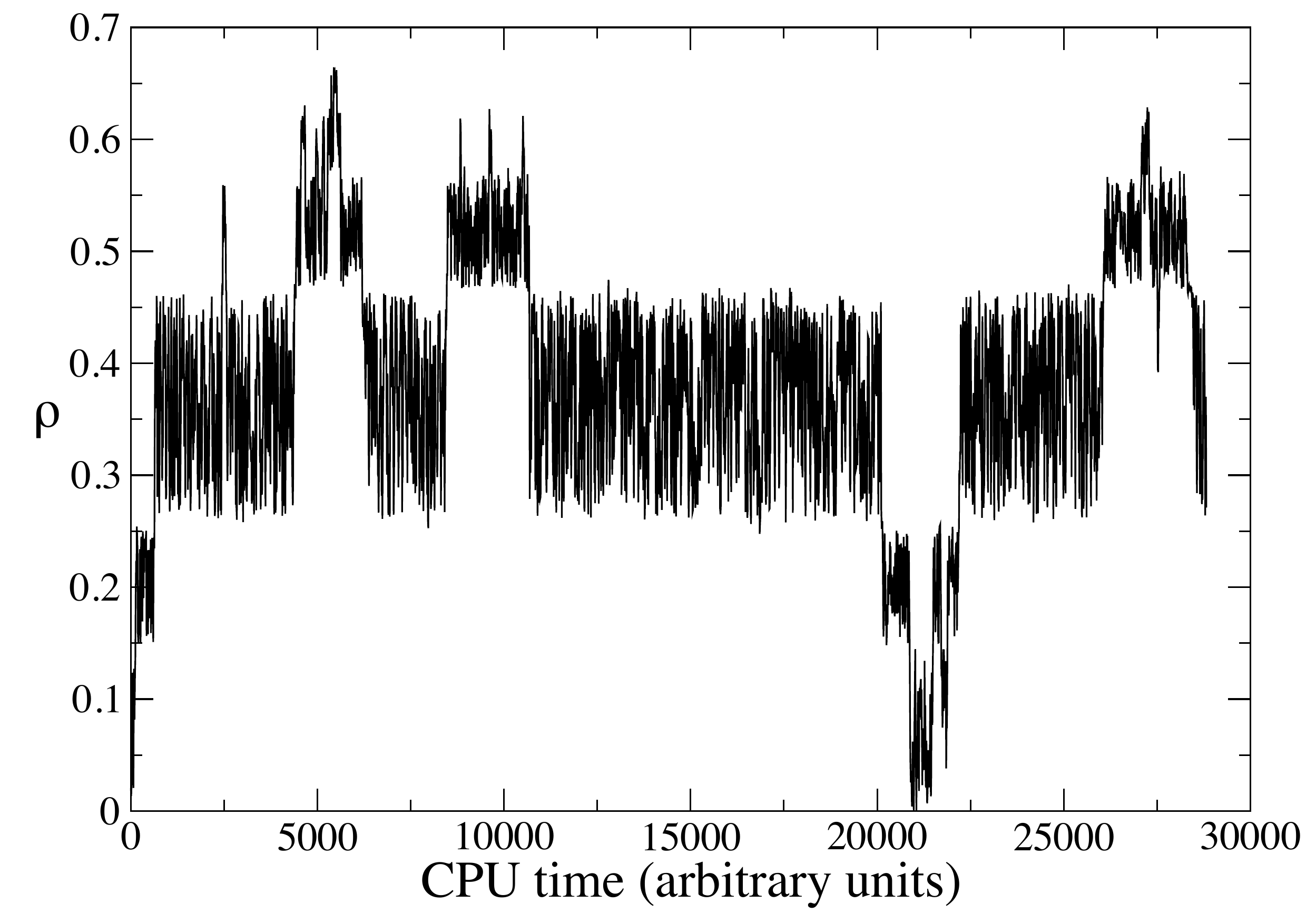}
\caption{A trajectory of a MUCA simulation of the LJ fluid at
  $T=0.775T_c$ biased with respect to $\rho$. The volume is
  $V=(20\sigma)^3$.  Spheres occur in the density range $\rho\approx
  0.07-0.15$; cylinders for $\rho\approx 0.15-0.26$, and slabs for
  $\rho\approx 0.26-0.46$. At still higher densities cylindrical and
  spherical bubble states occur.}
 \label{fig:traj}
 \end{figure}

Droplet transitions in fluids are of interest in their own right both
for the information they provide concerning curvature corrections to
the surface tension and for their finite-size scaling properties
\cite{MacDowell:2004wj,MacDowell:2006a,Binder:2012ez,Zierenberg:2015er}. Signatures
of the droplet transitions are visible in $p(\rho)$ which exhibits
kinks and rounded discontinuities at the densities for which the
transitions occur \cite{MacDowell:2006a,Binder:2012ez}. The kink at
the cylinder-slab transition is particularly pronounced
(cf. fig.~\ref{fig:prho} for $\rho\approx 0.26$). Unfortunately, for
large systems sizes and low temperatures, droplet transitions cause
serious sampling difficulties. These were first studied in the context
of MUCA simulations of the 2d Ising model where a finite-size scaling
analysis \cite{Neuhaus:2003pt} found that they lead to slowing down
which is exponential in $\beta L^{d-1}$.  Similar problems arise in
fluids \cite{Fischer:2010qw} and are manifest in the fact that even if
one finds a weight function $\eta(\rho)$ that leads to uniform
sampling of the density domain, the sampling becomes `stuck' in
regions of density associated with one type of
droplet. Fig.~\ref{fig:traj} illustrates this problem for the LJ
fluid. The root cause is that the density $\rho$ does not discriminate
sufficiently between the various droplet states, so 
biasing with a weight function $\eta(\rho)$ fails as a means for traversing the
transitions.

A number of ways of tackling this problem have been considered
previously. One is to augment the sampling with transitions to higher
temperatures where the barriers between transitions are smaller, which can be achieved
using expanded ensembles or parallel tempering
\cite{Grzelak:2010ya,Neuhaus:2007lm}. Although this is reported to
help, it seems rather cumbersome and laborious, particularly for very
low temperatures.  Another approach is to modify the system geometry
in order to remove the periodic boundaries that cause droplet
transitions. This has been tried in the case of a 2D fluid by
simulating it on the surface of a sphere \cite{Fischer:2010qw}. Doing
so appears to eliminate sampling barriers, but the use of curved
non-Euclidean space leads to non-trivial finite-size effects which
seem to affect calculations of surface tension in particular. Moreover
the extension to three dimensional systems seems complicated because
it involves simulating the system on the surface of a hypersphere.

A different way to try to tackle the sampling issues associated with
droplet transitions is to recognize that MUCA is not restricted to
weight functions that are defined solely with respect to simple
macrovariables such as the density or energy; any function of the
configurational coordinates can be used. Thus, for instance, work has
been done to try to find a smooth path between liquid and solid phases
by biasing with respect to an order parameter which is related to the
degree of local crystalline order in the system \cite{Chopra:2006km}.
In the present contribution we look at two approaches based on
non-standard order parameters and apply them to the VL
transition of a LJ fluid. Our first order parameter 
comprises a linear combination of moments of the configuration shape
and is designed to create a sampling path between vapour and slab
states that suppresses spherical and cylindrical droplets. Our second
method is inspired by the insights gained from the first and simply
uses the density in a tetragonal subvolume of the system as the order
parameter. Both methods improve the sampling, though the second is
considerably faster and easier to implement. In the following we
describe the methods in turn and test their efficacy.

\section{Linear combination of moments}

We consider a GCMC simulation of an orthorhombic simulation cell of dimensions $L_x,L_y,L_z$ with periodic boundary conditions. For any configuration ${\bf r}^N$ one may calculate three geometrical second moments with respect to the orthogonal box axes 
\numparts
\bea
m_x &=& \sum_i^N(x_i-\bar{x})^2\\
m_y &=& \sum_i^N(y_i-\bar{y})^2\\
m_z &=& \sum_i^N(z_i-\bar{z})^2
\eea
\endnumparts
where ${\bf r_i}=(x_i,y_i,z_i) $ is the position vector of particle $i$ and $\bar{x},{\bar y},{\bar z}$ are the components of the centre of mass (COM).
Our first order parameter is based on these moments. To calculate them we need to determine the COM.  However, this is not trivial to calculate for a periodic system. The definition appropriate to infinite space ($\bar{x}=N^{-1}\sum_i^Nx_i$, {\em etc}.) clearly fails for a periodic system as is readily appreciated by considering the case of a single cluster which is split across the periodic boundary. To circumvent this problem, Bai and Breen \cite{Bai:2008qh} have introduced an efficient scheme which averages the coordinates in a higher dimensional space. The method works by mapping each dimension of the simulation box onto a unit circle in the two-dimensional (2D) plane and performing averages in this plane. Specifically, for the determination of ${\bar x}$ one calculates, for each particle $i$, an angle in the interval $(0,2\pi)$:

\be
\theta_i=2\pi(x_i/L_x)\:.
\ee
 The corresponding point on a unit circle has 2D coordinates
\numparts
\bea
\chi_i &=& \cos(\theta_i),\\
\psi_i &=& \sin(\theta_i)\:.
\eea
\endnumparts
One calculates the average of these coordinates for the full set of $\{\theta_i\}$ 
\numparts
\bea
\bar{\chi} &=& \frac{1}{N}\sum_i^N\chi_i,\\
\bar{\psi} &=& \frac{1}{N}\sum_i^N \psi_i\:.
\label{eq:chipsi}
\eea
\endnumparts
Then the average coordinates $\bar{\chi}$ and $\bar{\psi}$ are extended back onto the unit circle to generate a new angle 
\be
\bar \theta=\arctan2 (-\bar{\psi},-\bar{\chi})+\pi\:.
\ee
Finally, mapping this angle back on to the $x$-axis of the simulation box yields the $x$-component of the centre of mass as

\be
{\bar x}=(\bar \theta/2\pi) L_x\:.
\ee
Analogous calculations for the $\{y_i\}$ and  $\{z_i\}$ yield the ${\bar y}$ and ${\bar z}$ components of the COM.

While this calculation of the COM is more complex than the naive one,  changes to the COM in the GCE as particles are added or deleted can be calculated very efficiently because one need only keep track of the changes to the sums $\sum_i\chi_i$ and  $\sum_i\psi_i$ appearing in Eq.~\ref{eq:chipsi}. 

Unfortunately matters are not quite so straightforward with regards to calculating the moments  themselves. For an infinite system one can expand 

\be
m_x = \sum_i^N(x_i-\bar{x})^2= \sum_i x_i^2-2\bar{x}\sum_i x_i+N\bar{x}^2
\label{eq:expand}
\ee
and play a similar trick of calculating changes in the moments due to a shift in the centre of mass by keeping track of changes to the sums when we insert or delete particles. However for a finite periodic system we should apply the minimum image convention to the displacement $x_i-\bar{x}$ before we square it and sum.  Since the COM $\bar{x}$ wanders during the simulation this requires, in principle, a recalculation of the displacements following each insertion or deletion. This is a potentially expensive calculation, albeit one which is only $O(N)$ in complexity. To reduce the cost we divide our particles into two sets, those that are nearer the COM than some prespecified radius, and those that are outside this radius. For each insertion and deletion we only check particles outside the radius whether the minimum image convention needs to be applied. For those inside the radius we calculate the contribution to the moments immediately by exploiting the expansion Eq. $\ref{eq:expand}$ ie. by keeping track of the partial sums and the COM.

Having calculated the moments $m_x,m_y,m_z$ for a given configuration,
we order them according to magnitude from the smallest to the largest
and relabel them $a$, $b$ and $c$ respectively.  Clearly for a
configuration which is in a spherical droplet state one has on
symmetry grounds that $a\approx b\approx c$; for a cylinder one has
$a<b\approx c$, while for a slab $a \approx b < c$. We now define the
order parameter for our system as a linear combination of $a$,$b$ and
$c$. Specifically we write

\be
{\cal M}\equiv c-\beta b-\alpha a\:,
\label{eq:lcmiop}
\ee
where $\alpha>0, \beta>0$ and $\alpha+\beta<1$.  We refer to ${\cal
  M}$ as the linear combination of moments (LCM) order parameter. It gives rise to a one-dimensional weight function $\eta({\cal M})$.  The
rationale for the definition of this order parameter is to alter the
reaction coordinate in order to bypass the formation of spheres and
cylinders, but not slabs. To appreciate how this can work, consider
the case $\alpha=0, \beta=1$. For this choice spheres and cylinders
occur for ${\cal M}\approx 0$, but since the vapor state also has
${\cal M}\approx 0$ {\em and} has a lower free energy than spheres and
cylinders, these droplet states will be suppressed.

\begin{figure}[h]
    \centering                                                                               
\includegraphics[type=pdf,ext=.pdf,read=.pdf,width=0.8\columnwidth,clip=true]{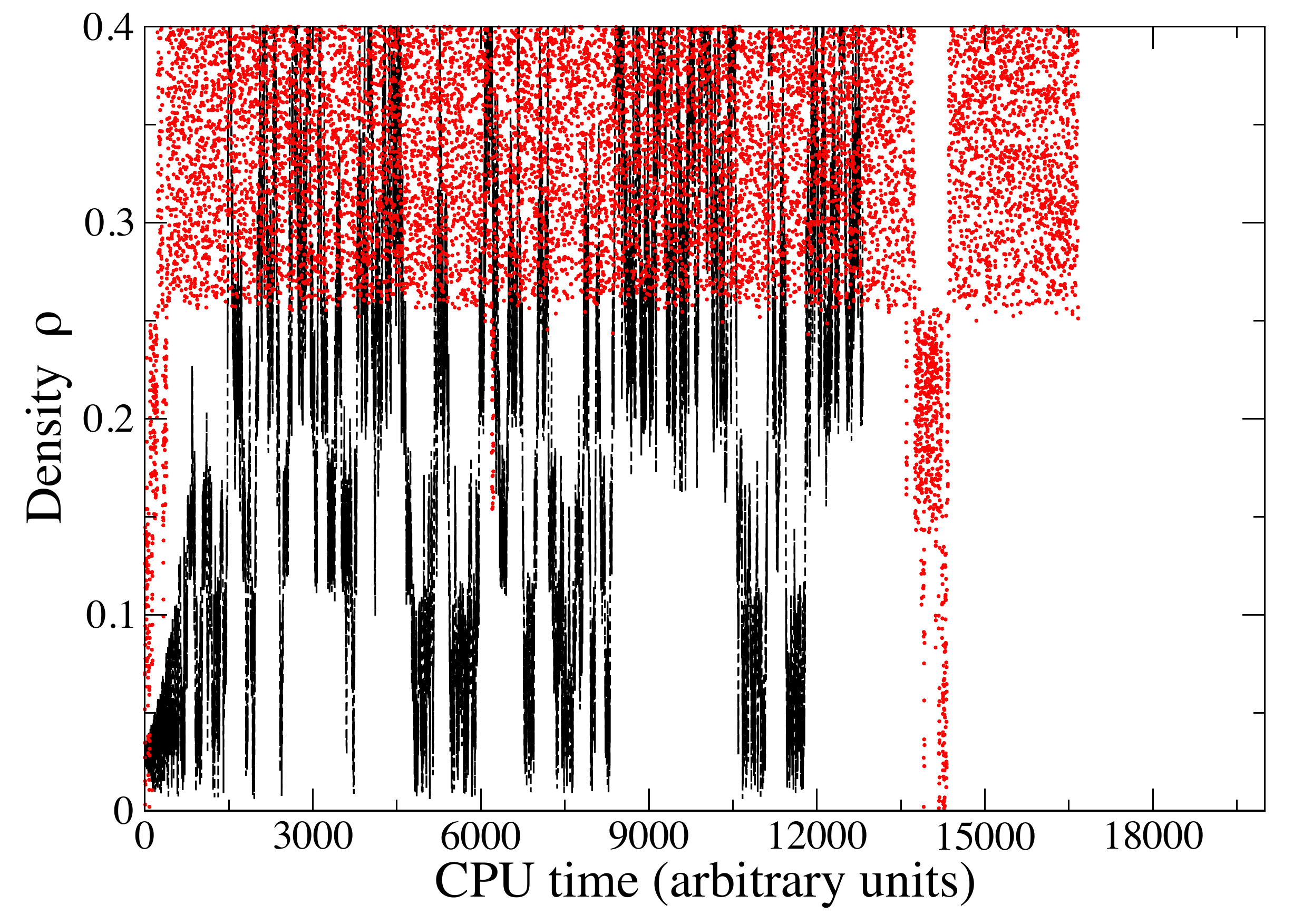}
\caption{Comparison of typical trajectories of the overall density $\rho$ as
  a function of CPU time (in arbitrary units) for the density order parameter (points)
  and the LCM order parameter (lines). The LCM parameters
  (cf. Eq.~\protect\ref{eq:lcmiop}) are $\beta=0.9,
  \alpha=0.1$. Clearly the barriers to sampling are much reduced for
  the LCM method. The system volume is $ V=(20\sigma)^3$ and the
  temperature is $T=0.775T_c$.}
 \label{fig:lcmicompare}
 \end{figure}

Clearly the LCM order parameter requires a greater computational
effort to calculate than a simple macrovariable such as the density or
energy.  Nevertheless, we find that this additional cost is more than
offset by improved sampling efficiency at vapour-liquid transitions.
In fig.~\ref{fig:lcmicompare} we compare the sampling efficiency of
MUCA simulations using the density order parameter and the LCM order parameter for a system of size
$L=20\sigma$, which is the largest for which one can observe droplet
state transitions on a reasonable time scale. The figure shows
the evolution of the density as a function of CPU time in arbitrary
units. For the LCM order
parameter many more round trips of the density range occur per unit
time than for the density order parameter. It is therefore more
efficient at finding the relative weight of vapor and slab states
which is necessary for accurate estimates of the surface tension.

\begin{figure}[h]
    \centering                                     
\includegraphics[type=pdf,ext=.pdf,read=.pdf,width=0.8\columnwidth,clip=true]{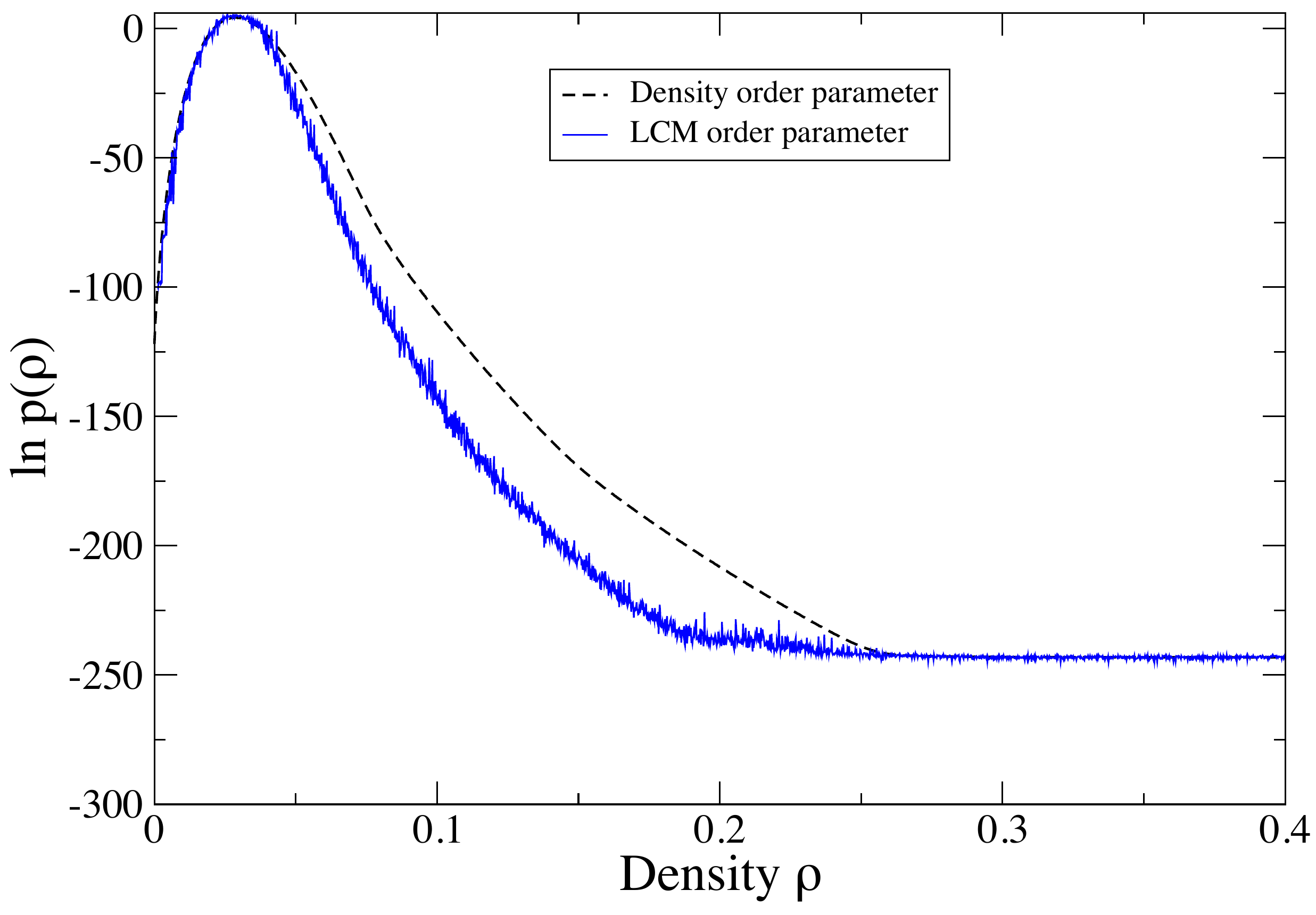}
\caption{Comparison of the logarithm of $p(\rho)$ determined from MUCA simulations biased with respect to the density order
  parameter and the LCM order parameter, as discussed in the text. The
  system volume is $ V=(17.5\sigma)^3$ and the temperature is
  $T=0.775T_c$.}
 \label{fig:dencomp}
 \end{figure}

The LCM order parameter and the density order parameter guide the
system along different configuration space paths. This becomes
apparent when one unfolds the weights from simulations performed using
the respective order parameters to obtain the forms of 
$p(\rho)$. These are compared in Fig.~\ref{fig:dencomp} for a system
of volume $V=(17.5\sigma)^3$ (which is the largest for which $p(\rho)$
can reliable be determined using the density
order parameter). The two distributions coincide in the region of the
vapor peak and in the (flat) region where slabs occur. However they
differ at intermediate densities because in the LCM case spheres and
cylinders are suppressed. Examination of configuration snapshots reveals
that for the LCM order parameter, the transition path favours
configurations which have slab-like moments from the outset. At small $\rho$, 
however, there are too few particles to form a true
slab and instead the system adopts a rather complex arrangement in
which two or more spherical droplets are arranged across the diagonal
of a notional slab.  In view of this it seem that the LCM method does
not eliminate droplet transitions altogether, although it does appear
to provide a significantly smoother sampling path from vapor to slab
than the traditional density order parameter.

\section{Biasing on a subvolume density}

We have seen that the LCM parameter engineers a smoother path between
vapor and liquid slab configurations by promoting configurations which
(at least in the sense of the relative magnitudes of the moments
describing their shape) are slab-like at all densities. This suggest a
simpler approach in which one increases the density from vapor-like
values by directly creating slab-like configurations. To this end we
implement MUCA with a bias defined on the number density
$\rho_s=N_s/V_s$ of particles in a tetragonal {\em subvolume} of dimensions
$V_s=L_x\times L_y\times d_z.$. Here the subvolume width $d_z$ is a
parameter which we set to lie in the range $5\sigma-7.5\sigma$. This
value is chosen to yield a slab that is sufficiently thick that it is
stable against capillary wave fluctuations of the VL interface, but
sufficiently narrow that spheres and cylinders cannot be comfortably
accommodated within the subvolume.

\begin{figure}[h]
    \centering                                   
\includegraphics[type=pdf,ext=.pdf,read=.pdf,width=0.8\columnwidth,clip=true]{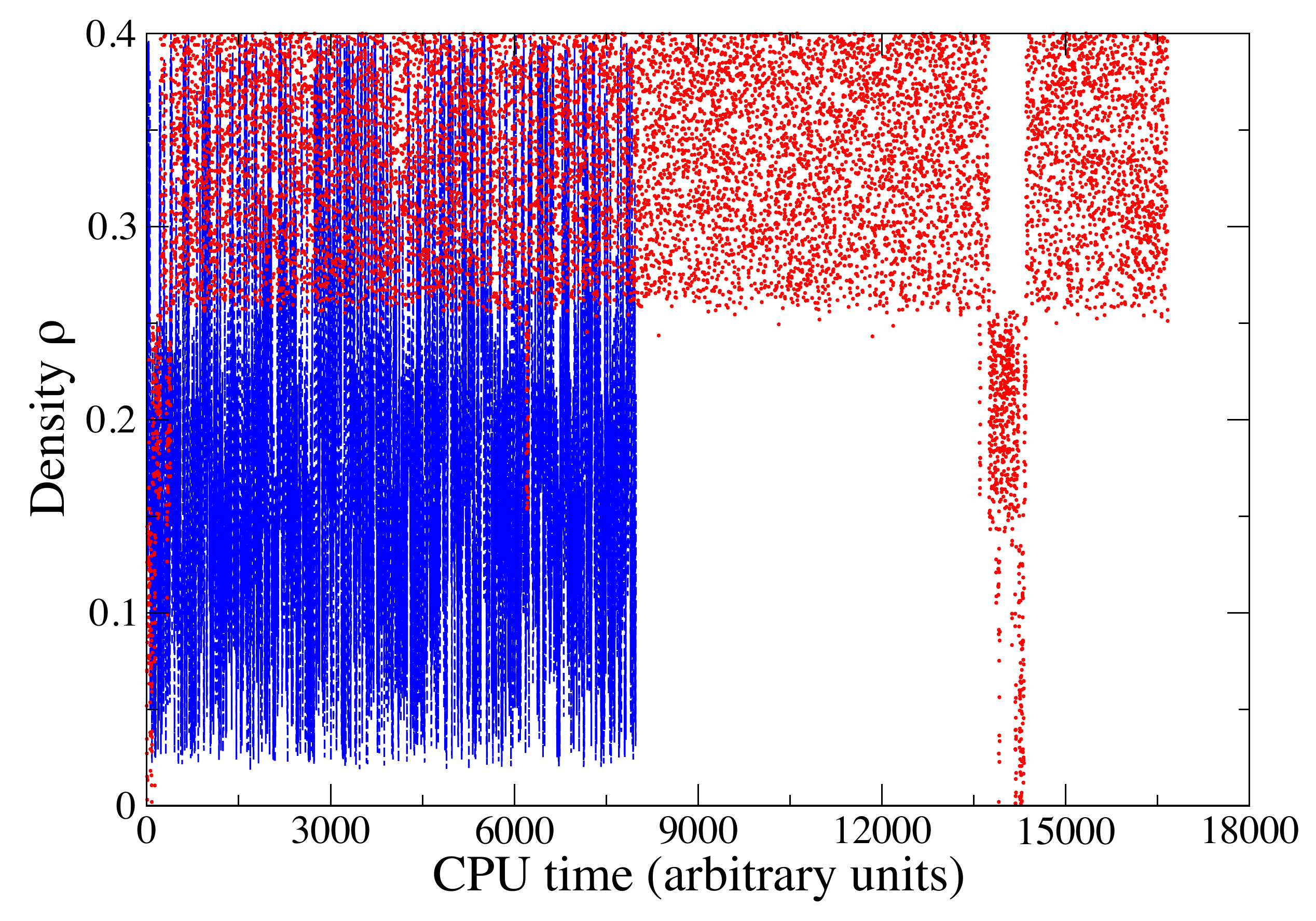}
\caption{Comparison of typical trajectories of the overall density as
  a function of time for MUCA simulations biased with respect to the
  overall density (points) and the subvolume density
  (lines). Clearly the sampling efficiency is greatly increased in the
  latter case. The system volume is $ V=(20\sigma)^3$ and the
  temperature is $T=0.775T_c$.}
 \label{fig:stripcomp}
 \end{figure}

By avoiding the costly moment calculation, simulations using the
subvolume density order parameter run about an order of magnitude
faster than the LCM approach. A comparison of sampling efficiency
presented in Fig.~\ref{fig:stripcomp} shows that the subvolume
approach reproduces the smoother sampling path between vapor and
liquid-slab densities seen in the LCM case, but the CPU time taken to
traverse this density range is only about $1\%$ of that required when
using the overall density order parameter. Use of the subvolume
approach therefore results in sampling which is about two orders of
magnitude faster than the standard density order parameter approach
for the system size studied. Like the LCM method, the subvolume
density order parameter method yields the same values of $p_{max}$ and
$p_{min}$ as does the overall density order parameter and can
therefore be used to calculate surface tension.  

The LCM method is based on geometrical moments of droplet
configurations which occur for densities between those of the pure
vapor and liquid slab configurations. At higher densities bubble
states appear which cannot be usefully characterised via the LCM order
parameter. By contrast the subvoume method can be used to traverse the
density range from vapor to slab and (separately) the density range
from slabs to pure liquid for which bubble transitions occur. To do so
one needs to calculate a separate MUCA weight function depending on
whether one is aiming to form a liquid slab within the vapor phase or
a vapor-like slab within the liquid phase. This ability to cover the
whole density range between vapor and liquid (albeit in two stages) is
a further reason to prefer the subvolume order parameter over the LCM
one.

\begin{figure}[h]
    \centering                               
\includegraphics[type=pdf,ext=.pdf,read=.pdf,width=0.8\columnwidth,clip=true]{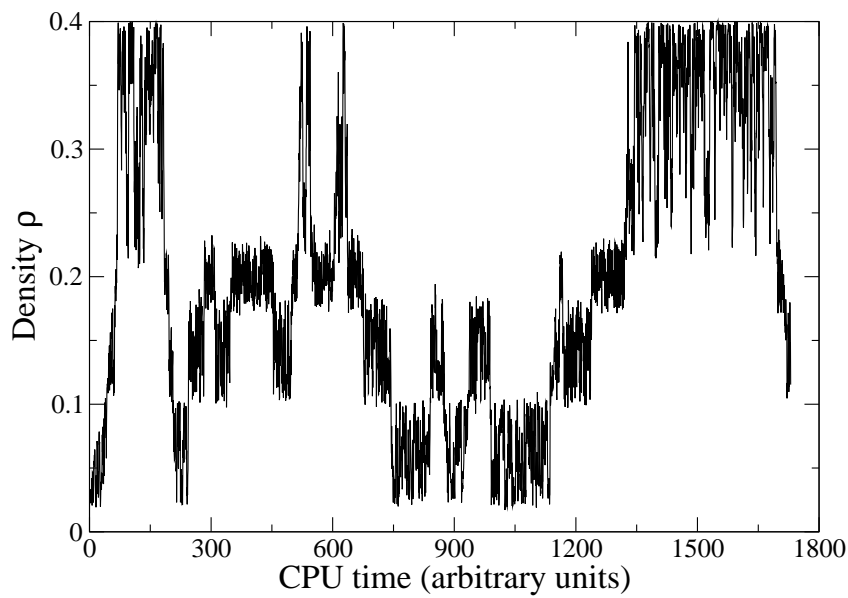}
\caption{A typical trajectory for the subvolume density method for $V=(25\sigma)^3$. Barriers to smooth sampling are clearly visible at this system size indicating the presence of droplet transitions.}
 \label{fig:10traj}
 \end{figure}

However, in common with the LCM case, the subvolume density
order parameter does not eliminate droplet transitions. This can be
seen by considering an even larger system size,
$V=(25\sigma)^3$. Examination of a typical trajectory
(Fig.~\ref{fig:10traj}) reveals the emergence of sampling barriers,
the scale of which we expect to grow strongly with increasing system
size.

\section{Conclusions}

In summary we have considered strategies for improving the efficiency
of multicanonical simulations of liquid vapor transitions at low
temperatures and large system sizes. Standard simulations based on
biasing with respect to the overall density suffer sampling
difficulties due to droplet transitions. We have considered two
alternative one-dimensional order parameters which ameliorate these
problems. One is based on a linear combination of the second moments
of the configuration geometry, the other is based on a subvolume
density. While both order parameters seem to reduce the height of
sampling barriers to a similar degree, the subvolume density method is
preferable because it is easier to implement, more versatile,  and entails little or no
computational overhead compared to biasing on the total density. We
expect that both order parameters will be applicable to lattice based
spin systems where droplet transitions also occur.

Although we have shown that alternative order parameters can reduce
the free energy barriers that cause slowing down, they clearly do not
eliminate them. The sphere-cylinder and cylinder-slab droplet
transitions are replaced by other transitions involving more complex
droplet geometries. It therefore remains the matter of further work to
devise better strategies that completely solve the sampling issues
that afflict GCMC at low temperatures and large systems sizes.

\ack

It is a pleasure to contribute to this special issue honouring the
achievements and memory of George Stell. The author recalls George's
kind hospitality on a visit to Stony Brook in 2002, and the stimulating
scientific discussions that led to a fruitful collaboration. George's
sharp intellect and imaginative approach, combined with a gentle,
humorous and supportive manner were all too rare. He will be much
missed.  \\[2mm]

\providecommand{\newblock}{}

\end{document}